\documentstyle[12pt]{article}

\begin{document} 
\begin{center}
{\bf \large Imitation of 2d quantum field theory by means of REM like models.}\\
\vspace{4mm}
D.B. Saakian\\

Yerevan Physics Institute,
Alikhanian Brothers St. 2,\\ Yerevan 375036, Armenia 

\end{center}

\begin{abstract}

An imitation of 2d field theory is formulated
by means of a model on the hierarchic tree (with branching number 
close to one) with the same 
potential and the free correlators identical to those of {\it 2d} ones. Such a model 
possesses some features of 
original models for certain scale invariant theories. For the case of 2d conformal
models it is possible to derive exact results. 
The renormalization group equation for the free energy is a reaction-diffusion equation,
 which  is noise-free KPZ equation with an additional linear term.  
For the case of Liouville model and strings these models on trees may be 
 naturally expressed via the Random Energy Model. This correspondence is used to identify 
the phase structure of strings for analytical continuation of DDK expressions. A phase
transition is found for spherical strings a bit below three dimensions.  

\end{abstract}

\section{Hierarchic tree with branching
number close to 1.}

One of the most fruitful ideas in physics is the idea of universality. 
In fact, that the only hope due to which rather artificial models
of the present theoretical physics can successfully capture the
relevant aspects of the nature.
We believe, that at the critical point the statistical 
mechanical system omits the secondary details. Usually this concerns the Hamiltonian in the $d$ 
dimensional Euclidean space.

It is proposed while keeping the Hamiltonian fixed to simplify the space geometry as much as possible 
retaining two point correlators and three 
point (for isosceles triangles) correlators. 
If the action of initial theory  consists of the Laplacian and a
potential, our model feels the space dimension through the behavior of Green
function
\begin{equation}
\label{e1}
G(x,x')\sim \frac{1}{r(x,x')^{d-2}}
\end{equation}
The total volume is 
\begin{equation}
\label{e2}
(\frac{L}{a})^d
\end{equation}
where {\it L} and {\it a} are infrared and ultraviolet cutoffs, $r(x,x')$ is the distance.
The Euclidean geometry has too many constructions. One can rotate a point around 
some center and circumscribe a close circle. 
Let us now consider some metric space with properties: \\
A.For every pair of points there is a distance r(x,x').\\
B.There is some measure at every point $d\mu_s(x)$ with the total measure $\int d \mu_s=R^d$.\\
C.One can construct a quadratic form with corresponding asymptotic (1) for the Green function.\\
We start out to construct statistical mechanical models on the 
simplest space, that supports points A-C.
We hope, that due to the universality these models will 
acquire some properties of models in d-dimensional space. 
To realize this program we will use some ideas from the theory
of Random Energy Model (REM) [1-5]. 
In ref. [5] a relation of 2d quantum Liouville model to REM 
and to the Directed Polymer (DP) on Cayley tree was established.  \\
Our present 
analysis shows, that the connection with REM is not a 
peculiar property of Liouville model and works well also for other 
conformal models. Besides,  
using similar ideas we intend to 
construct general 2d quantum models in the ultrametric space and thereby generalize the 
above-mentioned 
connection between the quantum field theoretical models and 
those defined on the hierarchical lattices.  
First, an ultrametric space (surface) with fields, located on the surface 
will be constructed. We define the following three geometrical objects:
the distance between surface points, the surface measure and the volume measure
for the ball, delimited by surface. These constructions 
are enough to define free field action with correlators that are identical
to those in the corresponding two-dimensional space. 
One can add also the interaction term to the action.
The practical merit of the proposed approach is that under certain conditions (for example, Coulomb gas approach to conformal
theories) the theories on the 
ultrametric space can be solved much easier, as compared to the 
Euclidean 2d space.\\
Let us define the ultrametric (UM) space with some measure for the
(spherical) area $d\mu_s(V,X)$, volume (ball) measure 
$d\mu_v(V,X)$, total surface $e^V$ and total volume
$e^V-1$. It is commonly known from the mean field  theory that the surface is of the same order 
of magnitude  as the volume.
We can construct this UM 
space as a limit of hierarchic lattices. Consider a tree with a constant
number of branching  q in each node  and number N for hierarchies. The 
number of end points is  $q^N$, the number of branches - 
$(q^{N}-1)/(q-1)$. We consider a set of end points as a surface of 
sphere, the set of branches making a volume (of the ball). Each point on 
the surface is connected with the origin  (zero level of the
hierarchy) via a single path, which consists of links.
We determine a measure $d\mu_s(V,X)=1$  for each end 
point, then $d\mu_l(V,l)=(q-1)$ for each link. This is mathematically 
correct in the limit $N\to \infty$.
Instead of integer
q  consider the limit $q\to 1, N\ln q\to V_0$. 
Now we have for the total area $\mu_s=\int d\mu_s(V_0,X)=e^{V_0}$
and for the total volume $\mu_v=\int d\mu_l(V_0,l)=e^{V_0}-1$. 
$d\mu_v\equiv\sum_l d\mu_l$, where the sum is over the links originated
from the point X. Later we will manipulate only with $d \mu_l$.
We choose $q\to 1$ to have an equation $\mu_l+1=\mu_s$.
We determine UM distance between two points x, y on this surface 
$V_0-v\equiv V_0-\frac{n}{N}$, where n is the number of the hierarchic 
level, on which $x$ and $y$ had the last common node on trajectories to their
point from the origin. The maximum UM distance between two points v on the
surface is $V_0$ (an ordinary distance as a function of V will be defined
lately). 
Now a scalar field $\phi(x)$ is defined on our surface.
For determination of kinetic energy, which should be a 
quadratic form with the
Laplacian as the kernel in the conventional space, let us consider the 
expansion 
\begin{equation}
\label{e3}
\phi(x)=f_0+\int_0^{1}{\it d}vf(v,l)
\end{equation}
Here $f(v,l)$ is determined on the links. The integration in (4) is made 
along the trajectory of point X. Since the measures on both the threads
of (4) coincide ($\int d\mu_l(V,l)=1+\int dV d\mu_l(V,l)$), the Jacobian
is equal to one.
Now determine the kinematics part of the action for the field $\phi(x)$ 
\begin{equation}
\label{e4}
\frac{1}{2}\int_0^{V_0}{\it d}V{\it d}\mu_l(V,l) f(V,l)^2
\end{equation}
Then the partition under the potential $U(\phi)$
\begin{eqnarray}
\label{e5}
\int {\it d}f\exp\{-\int_0^{V_0}{\it d}V\int{\it d}\mu_l(V,l)\frac{1}{2}
f(x,V)^2\}\nonumber\\
\exp\{\int d\mu_s(V_0,X)U(\phi(X))\}
\end{eqnarray}
We have for the correlator 
\begin{equation}
\label{e6}
<\phi(X)\phi(X')>=v
\end{equation}
where v is the UM distance between the points X, X'.
For usual 2d models with 
\begin{equation}
\label{e7}
\int {\it d}\phi_0{\it d}\phi\exp\{-\frac{1}{8\pi} {\it d}x^2\nabla \phi(x)^2\}
\exp\{\int d x U(\phi(x))\}
\end{equation}
the total surface area is equal to $R^2$, and the correlators read as
\begin{equation}
\label{e8}
<\phi(X)\phi(X')>=\ln \frac{L^2}{r^2}
\end{equation} 
In Eq. (7) it is possible to take n-component fields  instead of the one-component one $\phi(x)$.
We can determine the distance from the equality $V=\ln\ r^2$. Then our
correlators coincide (at any rate $r\gg 1$).
It is possible to construct a quantum field theory in this case. Our constructions for the
measure and distance are sufficient. One should bear in mind  only, that the volume measure
inside the sphere V is $e^V-1$.

We are going to discretize (5), then derive the iteration equations for imitations of 2d and 3d  
cases. For the case of spin models on hierarchic trees it is well known that it is 
possible to write simple iteration equations (similar to those in Ref.[6]) for any value of q. For the case (5) we formulated iteration equations in Section 2 for 2-d imitation and in Section 3  for 3-d case. The resulting equation are similar to the KPZ 
equation [7],but there is an additional linear term.  Instead of our abstract approach ($q=1$ trees) it is possible to consider a branching diffusion process like [4] and formulate 2-d models imitation on its basis. 
This approach is constructed in Section 4.

It is possible to derive the majority of results of 2-d conformal theory  by means of Coulomb gas representation [8]. Here the free field action is modified by an imaginary linear term in the action.
Correlators of fluctuating field in a critical theories with interaction are equivalent 
to correlators (from the exponential of a such free field) in modified free
field picture. In Section 5 we derived the three point correlators in our approach and the results agree with those in the 2-d case [8].

For the case of exponential function for $U(x)$  potential (2d Liouville model in the Euclidean space) there is a strict result [9]-[10] that the thermodynamics of our model on hierarchic tree
is independent of q and is similar to REM. In Section 6 a qualitative derivation 
for a REM with complex replica numbers (using the results of [11] for real number of replicas)is given and the solution of REM at complex temperatures [12]-[13] is obtained
 then those results are used to identify phase structure
of strings, using analytical continuation of DDK formulas [14]-[16]. In Section 6 the main results of the work are discussed.

\section{Iteration equations for the 2-d case.}

The advantage of 
representation (5) is that we are in a position to calculate the partition function
through iterations. This is well known for models on hierarchical
lattices [6].  Let us for some large number K, divide V into K
 parts $V/K$ and determine a hierarchical
 tree with K levels and branching number 
\begin{equation}
\label{e9}
q\equiv\exp\{\frac{V}{K}\}
\end{equation}

 Similar to [5] it is easy to define the partition Z via
 iterations. In case of some large number K we derive 
\begin{eqnarray}
\label{e10}
I_1(x)=\sqrt{\frac{K}{2\pi}}\int _{-\infty}^{\infty}\exp\{-\frac{1}{2}K y^2+U(x+y)\}{\it d }y\nonumber\\
I_{i+1}(x)=\sqrt{\frac{K}{2\pi}}\int _{-\infty}^{\infty}\exp\{-\frac{1}{2}K y^2\}
[I_{i}(x+y)]^{q}{\it d }y\nonumber\\
Z=\lim_{K\to \infty}[I_{K}(0)]^{q}
\end{eqnarray}
As for the determination of partition function, 
we need only the equation (10).
Our choice $q\to 1$ is a reasonable simplification. It is possible to construct perturbative 
field theory, calculate diagrams. To solve analytically equation (10) it is convenient
to consider the other (opposite) case, $q\to \infty$.  Only in this case  the analytical solution proves possible. The point is, that {\bf for the bulk structure of theory the value of q is irrelevant.}
 For example, in asymptotic expansion of free energy 
\begin{eqnarray}
\label{e11}
F(N)=F_0N+F_1\ln N+F_2\dots
\end{eqnarray}
only the last term depends on the choice of q (in case of Directed Polymer $N\sim V$).\\
We found how to construct the simplified version of any 2d theory on a hierarchical lattice.

Let us consider carefully equation (10) in the limit of large K and introduce  variable $w(v,x)\equiv I_{\frac{Kv}{V}}(x)$. We consider the limit $\frac{V}{K}\ll 1$. For the differential $dv$ we have an expression $\frac{V}{K}$. Let us also assume
\begin{eqnarray}
\label{e12}
q-1=\frac{V}{K}\equiv dv
\end{eqnarray}
Using the expression  $x^q\approx x(1+\log x(q-1))$ it is easy to obtain
\begin{eqnarray}
\label{e13}
\frac{d w}{d v}=w\ln w +\frac{1}{2}\Delta w\nonumber\\
w(0,x)=\exp(-U(x))
\end{eqnarray}
After the replacement  $w=\exp(u(t,x)$ we arrive at
\begin{eqnarray}
\label{e14}
\frac{d u}{d v}=\frac{1}{2}\Delta u+\frac{1}{2}(\nabla u)^2+u\nonumber\\
u(0,x)=U(x)
\end{eqnarray}
where $U(x)$ is the potential in Eq.(7). Having an expression for $u(v,x)$, we obtain 
for the free energy 
\begin{eqnarray}
\label{e15}
\ln Z=u(V,0)
\end{eqnarray}
We have a noise-free KPZ equation (14) with additional linear term for the free energy.\\
There are two interesting solution of Eq. (14) at large values v. If the couplings in the polynomial potential are $O(1)$,it is reasonable at large values of v  to consider the solution:
\begin{equation}
\label{e16}
u(v,x)=const \exp(v)
\end{equation}
If one considers the couplings $\sim \frac{1}{\exp(V)}$ in the potential $U(x)$ , then   the solution
\begin{equation}
\label{e17}
u(v,x)=const\exp(v)+u_s(x), u_s(x)\sim 1.
\end{equation} 

\begin{equation}
\label{e18}
\frac{1}{2}\Delta u_s+\frac{1}{2}(\nabla u_s)^2+u_s=0
\end{equation}
corresponds to the perturbative regime.
This equation gives the effective potential at the stable point of renormalization group.
One can rewrite of Eq. (18) in another form for $z\equiv \frac{d u_s}{dv}$:
\begin{equation}
\label{e19}
\frac{dz}{du_s}+z+\frac{2}{z}u_s
\end{equation}
In analogy to Eqs. (10),(13) it is possible also to derive the correlators. 
 To calculate the correlator 
$<\exp(i\alpha\phi(x)-i\alpha\phi(y))>$ , where the hierarchic distance between points $x,y$
is $v_0$,  one should distinguish during the iteration between the links located on the paths that connect the origin with the points x,y. Thus we also
consider the equation
\begin{eqnarray}
\label{e20}
\frac{d f(v,x,\alpha)}{d v}=f\ln w +\frac{1}{2}\Delta f\nonumber\\
f(0,x,\alpha)=\exp(U(x)+i\alpha x)
\end{eqnarray}
Then for the generating function $f_0(v,x)$ of correlator one must solve again Eq. (20) with the boundary conditions at the point $v_0$
\begin{eqnarray}
\label{e21}
f_0(v_0,x)=f(v_0,x,\alpha)f(v_0,x,-\alpha)/w(v_0,x)
\end{eqnarray}
We obtain an expression for the correlator :
\begin{eqnarray}
\label{e22}
<\exp(i\alpha\phi(x)-\alpha\phi(y))>=\frac{f_0(\infty,0)}{w(\infty,0)}
\end{eqnarray} 
In this way we can calculate two point correlators, as well as other multipoint correlators.

\section{High dimensions} 

The same approach may be used for the case of $d>2$.
In d-d space  one has for the volume $\sim a^d$. If we identify it with our $q^L$, then 
$a=q^{\frac{1}{d}L}$.
To have exact expression for the correlator, $f_l$ are defined on the branches,
$f_0,f_1$ at the origin. Here the free field action is defined as
\begin{equation}
\label{e23}
\phi(x)=f_0+f_1+\sum_{vl} f(v,l)
\end{equation}
The summation in (23) is along the trajectory of point X. 
Now determine the kinematical part of the action for $\phi(x)$  field 
\begin{equation}
\label{e24}
A=\frac{1}{2}[{f_1}^2+ \sum_{vl} \exp(-\alpha v) f(v,l)^2/\alpha]
\end{equation}
 If one takes 
$\alpha=\frac{d-2}{d}$ for the combined field,
then
\begin{eqnarray}
\label{e25}
<\phi(x)\phi(x')>=\exp(\alpha v) \sim [\frac{L}{r(x,x')}]^{-(d-2)}
\end{eqnarray}
where L is the infrared cutoff.
Now (10) transforms into:
\begin{eqnarray}
\label{e26}
I_1(x)=\sqrt{\frac{K}{2V\alpha \pi}}\int _{-\infty}^{\infty}\exp\{-\frac{K}{2V\alpha} y^2+U(x+y)\}{\it d }y\nonumber\\
I_{i+1}(x)=\sqrt{\frac{K}{2V\alpha e^{\alpha V(K-i+1)/K}\pi}}\int _{-\infty}^{\infty}e^{\{
-\frac{K}{2V\alpha \exp[\alpha V(K-i+1)/K]} y^2\}}
[I_{i}(x+y)]^q{\it d }y\nonumber\\
Z=\lim_{K\to \infty}\sqrt{\frac{1}{2\pi}}\int _{-\infty}^{\infty}\exp\{-\frac{1}{2} y^2+U(y)\}{\it d }y[I_{K}(y)]^{q}
\end{eqnarray}
To calculate $I_K(x)\equiv w(V,x)\equiv \exp(u(V,x))$ we have to solve the equation like (14)
\begin{eqnarray}
\label{e27}
\frac{d u}{d v}=\frac{1}{2}d(d-2) \exp[d(d-2) v]\Delta u+\frac{1}{2}(\nabla u)^2+u\nonumber\\
u(0,x)=U(x)
\end{eqnarray}
It is important to investigate the version of Generalized Random 
Energy Model (GREM) corresponding to (27) , when our model is defined on hierarchic lattice with large 
branching number q. Here the physics at 
 $d=2$ and $d>2$ is quite different and one cannot use the methods of [9] for the latter case.\\
\section{Branching diffusion}

Instead of the  structure of $q=1$ trees it is possible to consider an ensemble of hierarchic trees having again  a small branching number after the averaging. An analogous process has been introduced in [4], and here we give only a little modification to
 choose proper boundary conditions for
a desired potential.\\
There is an origin and a branch from it. Branch appears from the original branchs during the period of time {\it dt} with the length {\it dt}. Alternatively, with the probability 
{\it 1-dt} the old branch is elongated 
by the length {\it dt} and    a  random variable $f_l$ is introduced with the variance
\begin{equation}
\label{e28}
<f_l^2>=dt/2
\end{equation}
 After some period of time {\it t } the number of endpoints is
$\exp(t)$. Every branch has one or more random variables. The fields  at the endpoints of tree are again defined  
as a sum of random variables along the trajectory. The variable $w(t,x)$ is defined as
\begin{equation}
\label{e29}
w(t,x)=<exp(\sum_yU(y+x))>
\end{equation}
Here the summation {\it y} is along all endpoints of a tree after period of time {\it t}.
Let us define $w(t+dt,x)$. 
One should keep in view at the determination of $w(t+dt,x)$ that during {\it dt} time either there appears a new branch (the contribution of this process is equal to $dtw(t,x)^2$), or , if stays the old branch has an increment of $f_l$, the  contribution 
is $(1-dt)w(t,x+f_l)$. Combining the contributions we easily derive the KPP equation [9] for a {\it w} :
\begin{eqnarray}
\label{e30}
\frac{d w(t,x)}{d t}=w^2(t,x)-w(t,x) +\frac{1}{2}\Delta w
\end{eqnarray}
The derivation of this equation has been done in [9] where the case $w(0,x)=\exp(-e^x)$
has been considered. 
It is similar to (13), with the  only difference that the nonlinear term is replaced by $w^2-w$. We believe that
critical properties of both the approaches  (13),(30) are the same.
The point is that for $q=1$ tree approach there is no any averaging in an ensemble and thus
it is easier to deal with the perturbative expansion.
Of course our equation (13) is also a reaction diffusion-equation. \\
Another interesting problem is an imitation of d-d reaction diffusion equations
on $q=1$ space.

\section{Coulomb gas representation for 2d conformal fields.}
We can apply these ideas to conformal theories using the 
Coulomb-gas formalism with the  background charge $\alpha_0$ [8]. If we correctly
 defined the zero mode of Laplacian and the correlator has correct dimension, we have good chances
  to imitate the 2d situation. We have an action 
\begin{eqnarray}
\label{e31}
{\frac
{1}{8\pi}\int d^2w\sqrt{\hat g}{\phi \Delta \phi +i2\sqrt{2}\alpha_0R\phi}}
\end{eqnarray}
Here the field $\phi(w)$ is defined on the sphere, R is curvature, $\Delta$ is a Laplacian.
One defines the screening charges from the condition $\alpha_{\pm 1}^2-2\alpha_0\alpha_{\pm 1}=1$.
To calculate the correlator
$\prod_k\exp\{i\sqrt{2}\alpha_k\phi\}$
with
screening charges $Q_+^mQ_-^n$ one has to consider
\begin{eqnarray}
\label{e32}
Z=\int D_g\phi e^{\frac
{1}{8\pi}\int d^2w\sqrt{\hat g}{\phi \Delta \phi
+i2\sqrt{2}\alpha_0R\phi}}\nonumber\\
\prod_k\exp\{i\sqrt{2}\alpha_k\phi\}
(\int d^2 w\exp(i\sqrt{2}\alpha_+\phi)^n
(\int d^2 w\exp(i\sqrt{2}\alpha_-\phi)^m
\end{eqnarray}
The zero mode integration gives for nonzero correlator
the constraint 
\begin{eqnarray}
\label{e33}
\sum_i\alpha_i+m\alpha_++n\alpha_-=2\alpha_0
\end{eqnarray}
 in 2d case [8]. \\
To have finite set of $\alpha_i$ from (35) we should put a constraint 
\begin{eqnarray}
\label{e34}
p'\alpha_++p\alpha_-=0
\end{eqnarray}
which is the definition of minimal models. The deficiency of our approach is that 
we cannot find connection between $\alpha_0$ and conformal charge c.\\
While calculating (26) in UM space, we omit the $\alpha_0$ term.
We again , as in 2d case, consider normal ordered operator product for 
$\exp[i\sqrt{2}\alpha E_i]$
\begin{eqnarray}
\label{e35}
<z^{m}(\sqrt{2} \alpha_+)z^{n}(\sqrt{2} \alpha_-)\exp[i\sqrt{2} \alpha E_1] \exp[i\sqrt{2} \alpha E_2]>
\end{eqnarray} 
Here the average is over the normal distribution on our hierarchic tree. As m and n are integers,
 it is possible to perform the integration via $E_i$ directly. \\
How one can derive the expression for the pair correlator? All the $n+m$ charges with the 
total charge $2\alpha_0-\alpha$ are located near the point 1, or the point 2, so  we have for (35)
\begin{eqnarray}
\label{e36}
<z^{m}(\sqrt{2} \alpha_+)z^{n}(\sqrt{2} \alpha_-)\exp[i\sqrt{2} \alpha E_1] \exp[i\sqrt{2} \alpha E_2]>\sim\nonumber\\
\exp[\alpha(2\alpha_0-\alpha)v]=\exp[\alpha(2\alpha_0-\alpha)\ln \frac{L^2}{r^2}]
\end{eqnarray} 
For 3 point correlators we now consider  expressions like 
\begin{eqnarray}
\label{e37}
<z^{m}(\sqrt{2} \alpha_+)z^{n}(\sqrt{2} \alpha_-)\exp[i\sqrt{2} \alpha_1E_1] \exp[i\sqrt{2} \alpha_2E_2]
\exp[i\sqrt{2} \alpha_3E_3]>\nonumber\\
v_{13}=v_{23}>v{12}=v
\end{eqnarray} 
We assume that the screening charges are near the points 1 and 2. Then the distance from the 
point 3 to any charge is equal to  R and we immediately obtain for the dependence of (37) on R:
\begin{eqnarray}
\label{e38}
R^{-2\alpha_3(2\alpha_0-\alpha_3)}
\end{eqnarray} 
Let us consider the dependence of (32) on $r=L\exp[\frac{v}{2}]$.\\
As a result of direct interaction of charges $\alpha_1,\alpha_2$ we obtain:
\begin{eqnarray}
\label{e39}
r^{2\alpha_1\alpha_2}
\end{eqnarray} 
If all our $n+m$ charges are located at the distance r from both the points $1,2$, then their  
interaction energy with charges $1,2$ is proportional to
\begin{eqnarray}
\label{e40}
2(\alpha_1+\alpha_2)(n\alpha_++m\alpha_-)
\end{eqnarray} 
For the energy of self-interaction of the screening charges we have:
\begin{eqnarray}
\label{e41}
[(n\alpha_++m\alpha_-)^2-n\alpha_+^2-m\alpha_-^2]\nonumber\\
=(n\alpha_++m\alpha_-)^2-(n+m)-2\alpha_0(n\alpha_++m\alpha_-)]
\end{eqnarray} 
The integration over the coordinates of $n+m$ charges gives $\exp[3v]$, so eventually we have:
\begin{eqnarray}
\label{e42}
r^{-\alpha_3(2\alpha_0-\alpha_3)}r^{2\alpha_1\alpha_2+(n\alpha_++m\alpha_-)^2-2\alpha_0(n\alpha_++m\alpha_-)}
\end{eqnarray} 
It is easy to check that this expression is equivalent to standard expression
from the conformal field theory 
\begin{eqnarray}
\label{e43}
r^{-2\alpha_3(2\alpha_0-\alpha_3)}r^{\alpha_1^2-2\alpha_1\alpha_0+\alpha_2^2-2\alpha_2\alpha_0+
-\alpha_3^2+2\alpha_3\alpha_0}
\end{eqnarray} 
Combining two expressions, we derive eventually for the correlator:
\begin{eqnarray}
\label{e44}
R^{-2\alpha_3(2\alpha_0-\alpha_3)}r^{2\alpha_1\alpha_2+(n\alpha_++m\alpha_-)^2-2\alpha_0(n\alpha_++m\alpha_-)}
\end{eqnarray} 
Let us put in [43] $r_{12}=r_{13}=r_{23}$ in (43) and consider REM instead of the Directed Polymer. In case of REM
we have that $<\phi(X)\phi(X')>=\delta(x-x')$.  It is possible to investigate 
the phase structure of (originally DP) correlator in this way.

\section{REM at complex temperatures with complex numbers of replicas and strings.}
\subsection{REM version of strings}
Using similar ideas we will connect a string partition (after integration by zero mode)
with finite replica REM and investigate the phase structure. \\
Recall some results from string theory for a string in $d=c$ space with spherical surfaces [14-16]. It is known for the partition  that after integration by zero mode
\begin{eqnarray}
\label{e45}
Z\sim
\int D_g\phi e^{\frac
{1}{8\pi}\int d^2w\sqrt{\hat g}{\phi \Delta \phi +QR\phi}}
(\int d^2w 
\sqrt{\hat g}e^{\alpha\phi})^{-\frac{Q}{\alpha}}
\end{eqnarray}
where
\begin{eqnarray}
\label{e46}
c=1-12\alpha_0^2,
Q=2\sqrt{2+\alpha_0^2},
\alpha=-\frac{\sqrt{25-c}}{\sqrt{12}}+\frac{\sqrt{1-c}}{\sqrt{12}}\nonumber\\
\frac{Q}{\alpha}=\frac{1}{12}[c-25-\sqrt{(25-c)(1-c)}]\nonumber\\
\end{eqnarray}
Here $\phi(w)$ is a field on two dimensional sphere with coordinates {\it w},
curvature R.
If we continue those formulas for $c>1$, the coefficients become complex. \\
We see, that $ Z \sim <z^{\mu}>$, where  the averaging is over the normal
distribution of field $\phi(w)$.\\
If the conjecture about the equivalence of 2-d model (49) and corresponding $q=1$ model 
is correct (it could be checked numerically), then one can solve explicitly
DP problem in $Q\to \infty$ limit [9]-[10]. In this limit it is easy to prove [10], 
that the Directed Polymer has the same thermodynamic limit ($F_0$), as a simple REM. \\
If in (45) our variables  $\phi (w)$ are distributed according to the normal law $e^{\frac
{1}{8\pi}\int d^2w\sqrt{\hat g}{\phi \Delta \phi +QR\phi}}$ with non-diagonal quadratic form,
in the case of REM all $\phi(w)$ are independent variables with a normal distribution.
If we replace the model (45),(46) with models on $q=1$ trees, then according to [9]-[10] these models are 
equivalent to REM. Thus it is worthwhile to consider REM instead of the set (45)-(46).\\
To construct an equivalent scheme of REM let us introduce infrared and ultraviolet cutoffs L 
and a. Then the physical number of degrees is
\begin{eqnarray}
\label{e47}
M=\frac{L^2}{a^2}
\end{eqnarray}
Now define the distribution of $\phi(w)$ over all points w, using the free field action from (45):
\begin{eqnarray}
\label{e48}
\rho(\phi_0)\equiv <\delta(\phi_0-\phi(w)>_{\phi(w)} \sim
\exp(-\frac{\phi_0^2}{2G(0)})
\end{eqnarray}
where  the    averaging is over the  distribution 
\begin{eqnarray}
\label{e49}
\rho(\phi(w))\sim e^{\frac
{1}{8\pi}\int d^2w\sqrt{\hat g}{\phi \Delta \phi +QR\phi}}
\end{eqnarray}
  and
\begin{eqnarray}
\label{e50}
G(0)=2\ln \frac{L}{a}
\end{eqnarray}
We can replace our system with a collection of M independent variables $E_i\sim \phi(w)$
with the distribution (43) instead of (44).
Our goal is to calculate
\begin{eqnarray}
\label{e51}
Z=<z^{\mu_1+i\mu_2}>,
z=\sum_ie^{-(\beta_1+i\beta_2)E_i}
\end{eqnarray}
It is possible to solve the system rigorously. Here we are giving a qualitative derivation,
(checked by our exact calculations). \\
Note that
\begin{eqnarray}
\label{e52}
N=G(0),\beta_c=\sqrt{\frac{2\ln M}{G(0)}}
\end{eqnarray}

\subsection{Solution of REM at complex replica numbers}
Let us consider Eq. (51)
 for positive integer values of $\mu$, 
where the averaging is made over the distribution (52) for each $E_i$. \\
There are only two competing terms in the sum and  two corresponding phases.
The first one is paramagnetic  (PM) phase , originated from the 
cross terms in $z^{\mu}$ expansion 
\begin{eqnarray}
\label{e53}
Z=
M^{\mu}<e^{-\beta E_{i_1}-\beta E_{i_2}-..\beta E_{i_{\mu}}}>\nonumber\\
\ln Z=\mu\ln M+N\frac{\beta^2\mu}{2}=N\frac{(\beta_c^2+\beta^2)\mu}{2}
\end{eqnarray}
The second one, the correlated paramagnetic (CPM)[11] is originated from the diagonal terms in (53) like
$e^{-\beta{\mu} E_i}$  
\begin{eqnarray}
\label{e54}
Z=<(\sum_{i=1}^Me^{-{\mu}\beta E_i})>\nonumber\\
\ln Z=\ln M+\frac{N\beta^2\mu^2}{2}=\frac{N(\beta_c^2+\beta^2\mu^2)}{2}
\end{eqnarray}
Let us consider the continuation of (54) at $\mu<1$. At critical $\beta_c$
its  entropy $\ln Z-\beta\frac{d \ln Z}{d \beta}$ disappears. Let as assume for $\ln Z$
in this region an expression proportional to $\beta$ (it is natural for a system with zero entropy) and $\mu$. The continuity of $\ln Z$ gives for the spin-glass (SG) phase 
\begin{eqnarray}
\label{e55}
\ln Z=N\mu\beta_c\beta
\end{eqnarray}
If we pass to complex temperatures [14]-[15], then (51) transforms to (it is easy to check this directly for integer $\mu$)
\begin{eqnarray}
\label{e56}
\ln Z=N\frac{(\beta_c^2+\beta_1^2-\beta_2^2)\mu}{2}
\end{eqnarray}
In Eq. (55) one has to replace $\beta$ by $\beta_1$,then
\begin{eqnarray}
\label{e57}
\ln Z=N\mu\beta_c\beta_1
\end{eqnarray}
For complex temperatures there is the fourth,  Lee-Young-Fisher (LYF) phase [12]. Its derivation is not direct. The point is that for noninteger values of $\mu$
\begin{eqnarray}
\label{e58}
Z\sim <|z|^{\mu}>
\end{eqnarray}
After this trick it is easy to derive the CPM expression. The principal terms are now 
given by terms
$e^{-2\beta_1 E_i}$
\begin{eqnarray}
\label{e59}
\ln Z=\frac{N(\beta_c^2+8\beta_1^2)\mu}{4}
\end{eqnarray}
Now continue our four expressions to complex values of $\mu$.\\
For PM phase
\begin{eqnarray}
\label{e60}
\ln Z=N\frac{(\beta_c^2+\beta_1^2-\beta_2^2)\mu_1}{2}
\end{eqnarray}
For SG phase
\begin{eqnarray}
\label{e61}
\ln Z=N\mu_1\beta_c\beta_1
\end{eqnarray}
For LYF phase
\begin{eqnarray}
\label{e62}
\ln Z=\frac{N(\beta_c^2+8\beta_1^2)\mu_1}{4}
\end{eqnarray}
For CPM
\begin{eqnarray}
\label{e63}
\ln Z=\frac{N[\beta_c^2+\beta_1^2(\mu_1^2-\mu_2^2)]}{2}
\end{eqnarray}
The imaginary parts of $\ln Z$ in (61)-(63) were ignored.\\
To find the borders between four phases first has to obtain  
the correct phase at $\mu\to 0$ limit, then compare its expression
for $|\ln Z|$ with the corresponding one for CPM phase. It is known that LYF phase 
exists at 
\begin{eqnarray}
\label{e64}
\beta<\frac{\beta_c}{2}
\end{eqnarray}
 and PM one at
$\beta<\beta_c$.
For complex temperatures one has a condition for SG phase
\begin{eqnarray}
\label{e65}
\beta_1>\beta_c+\beta_2
\end{eqnarray}
The last point. The rigorous derivation gives that LYF for noninteger $\mu_1$ exists only at
 \begin{eqnarray}
\label{e66}
\mu_1>-2
\end{eqnarray}

\subsection{REM results for the phase structure of string.}

One can apply this (though qualitative but rather strict) result to strings. Identifying
 $-\frac{\alpha}{\sqrt{2}}\to \beta, \frac{Q}{\alpha}\to \mu$ and using
(2) we derive
\begin{eqnarray}
\label{e67}
\beta_1=\frac{\sqrt{25-c}}{\sqrt{24}},\beta_2=\frac{\sqrt{c-1}}{\sqrt{24}}
,\mu_1=\frac{1}{12}[25-c], \mu_2=\sqrt{(25-c)(c-1)}\frac{1}{12}
\end{eqnarray}
For other string topologies one has
\begin{eqnarray}
\label{e68}
\mu\to (1-g)\mu
\end{eqnarray}
Note that $y=\frac{25-c}{24}$.
LYF phase exists only in the torus case at  $25>c>19$. At $19>c>1$ the system with torus topology and higher is in SG phase. For
sphere topology $19>c>1$, it is in CPM at
\begin{eqnarray}
\label{e69}
1+4y(2y-1)^2>4y^{\frac{1}{2}},
\end{eqnarray}
otherwise in SG phase.
We see phase transition at $d\approx 2.98$.\\
For $25>c>19$ spherical topology case CPM is at 
\begin{eqnarray}
\label{e70}
1+4y(2y-1)^2>\frac{1}{2}+4y^{2},
\end{eqnarray}
otherwise at LYF phase.
What can we say about string physics base on the REM picture? \\
The most interesting case is the spherical one. When the value of {\it c}is increased to pass  
 over the $ c=1$ barrier, nothing happens in REM picture, the system is still in CPM phase, as for in the $c<1$ case. 
The PM or CPM phases are ordinary physical phases, so one could try to succeed here with the same level of reliability,
as for $c<1$ strings outside of the minimal series. To reveal interesting (unitary) theories explicitly one should solve the directed polymer at finite replica number including finite size corrections and 
correlators.\\
For the high topologies we have omitted modular space dependence of partition. It will be interesting to check our conclusion about different physical phases for different
topologies numerically. But at least for spherical case the REM analysis seems quite reliable. 

\section{Conclusions}
We generalized the results of [5], obtained for the 2d model with an exponential potential, for the case of other critical models, that may be formulated by means of Laplasian (as a single 
differential operator)and a desired potential of any dimensions $d\ge 2$. The models can be formulated on hierarchic trees with constant branching number q. Main observation of this work is that for the special case of $q\to 1$ it is possible to construct
 a field theory that is
similar to renormalized field theories in continuous spaces.

We hope, that the bulk structure, the two and three point correlators (for isosceles
 triangles) are the same, as those in d-d critical models. This hypothesis is correct for the case of Liouville 2d model , as well as for the (free) case of Coulomb gas representation [8].

We have also analyzed the phase structure of strings by consideration of an analytical continuation 
of DDK formulas by analogy with REM and solution of complex replica REM at complex temperatures.

 It is  possible to check our hypothesis about the
 equivalence of our models on $q=1$ trees with some segment of d-d field theory by means of 
 direct numerical calculation of Eq. (14),(27), for example for, the field version of 3d Ising model 
 with proper choice of potential U. \\
I am grateful to  ISTC grant A-102 
 for partial financial support, C. Lang and W. Janke for invitation to Graz and Leipzig
and discussions, P. Grassberger for useful remark. 

\end{document}